\documentclass[twocolumn,showpacs,preprintnumbers,amsmath,amssymb]{revtex4}

\usepackage{graphicx}
\usepackage{dcolumn}
\usepackage{bm}

\begin{document}

\title{Unimodular Relativity and the Dark Matter Problem}
\author{Robert D. Bock\footnote{email: robert.bock@pra-corp.com}}

\affiliation{Propagation Research Associates, Inc., 1275 Kennestone Circle, Suite 100, Marietta, GA, 30066}

\date{\today}

\begin{abstract}
We introduce a modified divergence law for the energy-momentum tensor in the theory of unimodular relativity.  Consequently, an additional equation for the scalar curvature follows from the divergence of the field equations.  The equations of motion are derived and the weak-field, low-velocity limit is investigated.  It is found that the gravitational acceleration acquires a term that is proportional to the gradient of the mass density.  We show that this term can provide the additional acceleration observed on astrophysical scales without the need for dark matter.   
\end{abstract}

\maketitle

\section{\label{sec:introduction}Introduction}
Einstein's theory of general relativity (GR) \cite{einstein1916} has enjoyed overwhelming success since its inception, predicting phenomena that have been observed in our solar system and beyond, including the perihelion precession of Mercury, relativistic effects in the Hulse-Taylor binary pulsar B1913+16, and the existence of black holes.  While GR reduces to Newtonian gravity in the weak-field, low-velocity limit, and hence agrees with observation for a wide range of phenomena, anomalous accelerations have been observed for decades.  First discovered by Zwicky in the 1930's \cite{zwicky1933,zwicky1937}, velocities on the galactic scale are much larger than those predicted by GR when the source of the gravitational field is taken to be the observed visible matter (see also, e.g., \cite{einastoetal} and \cite{rubinetal}).  Zwicky postulated, and it is now generally accepted, that a considerable amount of non-visible matter must be present in the extragalactic regime in order to provide the additional acceleration required to maintain these excessive velocities.  This non-visible matter is commonly called dark matter and is believed to resolve acceleration discrepancies observed in systems ranging from dwarf spheroidal galaxies with visible masses $\sim 10^7 M_{\odot}$ to clusters of galaxies with observed masses $\sim 10^{14} M_{\odot}$.  Furthermore, dark matter is believed to play a key role in structure formation of the universe and primordial nucleosynthesis, and is believed to significantly affect the anisotropy of the cosmic microwave background.  Excellent reviews of the dark matter problem are given in Refs. \cite{turner,silk}.

Despite thirty years of laboratory experiments and astronomical observation, dark matter has never been observed directly \cite{Bertoneetal}; its existence is only inferred indirectly due to its purported gravitational effects on visible matter.  Modifications of gravitational theory have been proposed \cite{finzi,tohline,sanders1984,sanders1986,goldman,kuhnandkruglyak} that may eliminate the need for dark matter, and perhaps the most well-known is the modified Newtonian dynamics (MOND) theory \cite{milgrom1983a,milgrom1983b,milgrom1983c}.  MOND is characterized by an acceleration scale and predicts departures from a Newtonian force law in the extragalactic regime where dynamical accelerations are small.  Recently, a relativistic generalization of MOND, Tensor-Vector-Scalar (TeVeS), was proposed  \cite{bekenstein2004} that resolves some of the earlier problems of the MOND theory.  However, TeVeS has not been experimentally confirmed. 

In the following, we show that the introduction of a modified divergence law for the energy-momentum tensor in the unimodular theory of gravity results in an additional field equation.  In the weak-field, low-velocity limit, the equations of motion acquire a term that is characterized by gradients in mass density.  It is shown that these gradients can provide the additional acceleration observed on the galactic scale without the need for dark matter.  The field equations reduce to GR when these gradients are negligible.  Note that modified divergence laws are also discussed in \cite{rastall,smalley,tiwari}.

\section{\label{sec:fieldequations}Field Equations}
According to the theory of unimodular relativity \cite{andersonfinkelstein}, the metric tensor is reducible under the general coordinate category into two nontrivial geometric objects: $g$, the determinant of the metric tensor, and $\gamma_{\mu\nu}$, the relative tensor $g_{\mu\nu}/(\sqrt{-g})^{1/2}$ of determinant $-1$.  The determinant determines entirely the measure structure of space–time, while the relative tensor alone determines the null-cone or causal structure.  Unimodular relativity assumes a background measure field $\sqrt{-g}=\sigma(x^\alpha)$ so that:
\begin{equation}
\label{metric}
g_{\mu\nu}=\sigma(x^\alpha)^{1/2}\gamma_{\mu\nu}.
\end{equation}
This condition is satisifed with the method of Lagrange undetermined multipliers in the action integral:
\begin{equation}
\label{action}
S=\int\left[(R + 2\kappa L_m)\sqrt{-g}+\lambda(\sqrt{-g}-\sigma)\right] d^4x,
\end{equation}   
where $L_m$ is the Lagrangian density for the matter fields, $\lambda$ is the Lagrange undetermined multiplier, and $\kappa=\frac{8\pi G}{c^4}$ (see also \cite{buchmullerdragon,ngdam,weinberg,bock}).  The curvature scalar $R=g^{\mu\nu}R_{\mu\nu}$ is constructed from the connections: 
\begin{equation}
\label{connection}
\Gamma^{\mu}_{\alpha\beta}=\frac{g^{\mu\lambda}}{2}\left(g_{\alpha\lambda,\beta}+g_{\beta\lambda,\alpha}-g_{\alpha\beta,\lambda} \right),
\end{equation}
where a comma denotes partial derivative.  Variation of $\lambda$ gives $\sqrt{-g}=\sigma(x^\alpha)$ and variation of the quantities $g_{\mu\nu}$ yields the field equations:
\begin{equation}
\label{fieldequation1}
R^{\mu\nu}-\frac{1}{2}Rg^{\mu\nu}-\frac{1}{2}\lambda g^{\mu\nu}=-\kappa T^{\mu\nu}.
\end{equation}
Taking the trace of Eq. (\ref{fieldequation1}) gives:
\begin{equation}
\label{lambda}
\lambda = -\frac{1}{2}(R-\kappa T),
\end{equation}
and substituting this back into the field equations produces:
\begin{equation}
\label{traceless}
R^{\mu\nu}-\frac{1}{4}Rg^{\mu\nu}=-\left( \kappa T^{\mu\nu}-\frac{1}{4}\kappa Tg^{\mu\nu}\right).
\end{equation}
Taking the covariant divergence of these field equations gives:
\begin{equation}
\label{firstidentity}
\kappa T^{\mu\nu}_{\;\; ;\nu}=-\frac{1}{4}g^{\mu\nu}\left(R- \kappa T \right)_{,\nu},
\end{equation}
where the contracted Bianchi identity ($G^{\mu\nu}_{\;\; ;\nu}=0$) was used.  In the common formulations of unimodular relativity \cite{andersonfinkelstein,buchmullerdragon,ngdam,weinberg}, the covariant divergence law is assumed:
\begin{equation}
\label{matterconservation}
T^{\mu\nu}_{\;\; ;\nu} = \frac{(\sqrt{-g} T^{\mu\nu})_{,\nu}}{\sqrt{-g}}+\Gamma^{\mu}_{\alpha\nu}T^{\alpha\nu}=0,
\end{equation}
so that Eqs. (\ref{lambda}) and (\ref{firstidentity}) give:
\begin{equation}
\label{constantcosmologicalconstant}
\frac{\partial \lambda}{\partial x^{\mu}}=0.
\end{equation}
Hence, the cosmological constant emerges as a constant of integration in the field equations.  This formulation has the attractive property that the contribution of vacuum fluctuations automatically cancels on the right hand side of Eq. (\ref{traceless}) \cite{weinberg}.  

The primary difference between unimodular relativity and standard GR is revealed by evaluating the covariant divergence of the stress-energy tensor in each theory.  In standard GR, the covariant divergence of the stress-energy tensor is fixed from the field equations:
\begin{equation}
\label{fieldequationsGR}
R^{\mu\nu}-\frac{1}{2}Rg^{\mu\nu}=- \kappa T^{\mu\nu}.
\end{equation}
Taking the covariant divergence of each side gives $T^{\mu\nu}_{\;\; ;\nu}=0$.  In unimodular relativity, as in standard GR, the covariant divergence  also follows from the field equations (cf. Eq. (\ref{firstidentity})).  However, in unimodular relativity, the divergence is a function of the scalar curvature $R$.  Since the unimodular field equations are traceless, the scalar curvature $R$ is undetermined, and therefore one may not evaluate Eq. (\ref{firstidentity}) without an additional condition.  Typically, this additional condition is taken over from standard GR by simply forcing the covariant divergence to vanish in unimodular relativity as well \cite{andersonfinkelstein,buchmullerdragon,ngdam,weinberg}.  However, this is an unjustified assumption in unimodular relativity, which is highlighted by the fact that the measure field is non-dynamical and therefore unimodular relativity is fundamentally not invariant under arbitrary four-dimensional coordinate transformations.  Consequently, we cannot assume that the action for matter in unimodular relativity leads to the vanishing of the covariant divergence of the stress-energy tensor.  Of course one may write the unimodular action for matter in a diffeomorphism-invariant form; however, one cannot assume that in so doing this will lead to the standard divergence law of GR.    

Therefore, one must permit generalizations of the standard divergence law in unimodular relativity since $T^{\mu\nu}_{\;\; ;\nu}=0$ is an unjustified assumption.  One way to proceed is to re-examine the physical definition of divergence in light of the metric decomposition (\ref{metric}).  The covariant divergence, in turn, relies on the parallel transport (defined by the connections) of energy-momentum since the divergence calculation involves the sum of the fluxes of energy-momentum through infinitesimal hypersurfaces separated by infinitesimal space-time intervals.  Thus, we begin by examining the impact of the metric decomposition (\ref{metric}) on the connection (\ref{connection}).  Substituting Eq. (\ref{metric}) into Eq. (\ref{connection}) we find:
\begin{equation}
\label{connection2}
\Gamma^{\mu}_{\alpha\beta}=\Gamma^{\prime\mu}_{\alpha\beta}+\Gamma^{\prime \prime\mu}_{\alpha\beta},
\end{equation}
where 
\begin{eqnarray}
\label{connection2defs}
\Gamma^{\prime\mu}_{\alpha\beta} &\equiv & \frac{\gamma^{\mu\lambda}}{2}\left[\gamma_{\alpha \lambda ,\beta} + \gamma_{\beta \lambda ,\alpha} - \gamma_{\alpha \beta ,\lambda}\right] \\
\Gamma^{\prime \prime\mu}_{\alpha\beta} &\equiv &\frac{1}{4\sigma}\left[ \delta_{\alpha}^{\mu}\sigma_{,\beta} + \delta_{\beta}^{\mu}\sigma_{,\alpha} - \gamma^{\mu\lambda}\gamma_{\alpha\beta}\sigma_{,\lambda}\right],
\end{eqnarray}
and $\gamma^{\mu\nu}$ are the normalized minors of $\gamma_{\mu\nu}$.  We see that the connections separate into two terms, one that depends on derivatives of $\gamma_{\mu\nu}$, and another that depends on derivatives of $\sigma(x)$.  Hence, parallel transport may be decomposed into two fundamental entities:  parallel transport due to derivatives in the null-cone structure of space-time defined by $\Gamma^{\prime\mu}_{\alpha\beta}$, and parallel transport due to derivatives in the measure structure of space-time defined by $\Gamma^{\prime \prime\mu}_{\alpha\beta}$.  Substituting Eqs. (\ref{metric}) and (\ref{connection2}) into  $T^{\mu\nu}_{\;\; ;\nu}$ we obtain:
\begin{eqnarray}
\label{newdivergence}
T^{\mu\nu}_{\;\; ;\nu}=T^{\mu\nu}_{\;\; ||\nu}+\frac{\sigma_{,\nu}}{\sigma}
\left[\frac{3}{2}T^{\mu\nu}-\frac{1}{4} T g^{\mu\nu} \right],
\end{eqnarray}  
where 
\begin{equation}
\label{definenewdivergence}
T^{\mu\nu}_{\;\; ||\nu}\equiv T^{\mu\nu}_{\;\; ,\nu}+T^{\alpha\nu}\Gamma^{\prime\mu}_{\alpha\nu}+T^{\mu\nu}\Gamma^{\prime\alpha}_{\alpha\nu}.
\end{equation}
We see that the covariant divergence also separates naturally into two entities, one that looks like a divergence with respect to the null-cone structure alone, and another that depends on derivatives of the measure field.  Note that $\Gamma^{\prime\alpha}_{\alpha\nu}$ vanishes because $\sqrt{-g}_{;\lambda}=0$ and $\Gamma^{\alpha}_{\alpha\beta}=\Gamma^{\prime \alpha}_{\alpha\beta}+\frac{\sigma_{,\beta}}{\sigma}$.

As pointed out above, one may not adopt the additional condition $T^{\mu\nu}_{\;\; ;\nu}=0$ in unimodular relativity since diffeomorphism invariance has been relaxed and we do not know the transformation properties of the unimodular matter action a priori.  However, we may use the substructure of the connections and the covariant divergence identified above to postulate a generalized divergence law that is consistent with the relaxation of the invariance group.  In so doing, we will find a likely candidate for the matter Lagrangian in unimodular relativity (cf. Eqs. (\ref{lagrangian}) and (\ref{lagrangian2})).  We proceed by generalizing the law of parallel transport of energy-momentum so that the the null-cone and measure contributions may be weighted independently.  For example, consider a law of parallel transport of energy-momentum based only on the derivatives in the null-cone structure defined by the connection $\Gamma^{\prime \mu}_{\alpha\beta}$.  This leads to a divergence law based on the vanishing of Eq. (\ref{definenewdivergence}) alone:
\begin{equation}
\label{modconservation}
T^{\mu\nu}_{\;\; ||\nu}=0.
\end{equation}
In this case Eq. (\ref{matterconservation}) becomes:
\begin{equation}
\label{modconservation2}
T^{\mu\nu}_{\;\; ;\nu}=\frac{\sigma_{,\nu}}{\sigma}
\left[\frac{3}{2}T^{\mu\nu}-\frac{1}{4} T g^{\mu\nu} \right].
\end{equation}
More generally, we may consider parallel transport that is weighted arbitrarily by each term in Eq. (\ref{connection2}).  This leads to the following generalization of the divergence law:
\begin{equation}
\label{modconservation2b}
T^{\mu\nu}_{\;\; ;\nu}=\frac{\sigma_{,\nu}}{\sigma}
\left[AT^{\mu\nu}+B T g^{\mu\nu}\right] + C T^{\alpha\nu}\Gamma^{\prime\mu}_{\alpha\nu},
\end{equation}
where $A$, $B$, and $C$ are arbitrary constants.  This reduces to the standard covariant divergence law when $A=B=C=0$.  This is equivalent to Eq. (\ref{modconservation2}) for $A=3/2$, $B=-1/4$, and $C=0$.  This divergence law implies that the action for matter is not a scalar invariant.  When $A=C=0$ this divergence law leaves the usual covariant law unchanged for a traceless energy-momentum tensor (e.g., the electromagnetic energy-momentum tensor). 

The modified divergence law (\ref{modconservation2b}) can be used with Eq. (\ref{firstidentity}) to derive an equation for the scalar curvature:
\begin{equation}
\label{sigmaequationb}
\frac{\sigma_{,\nu}}{\sigma}
\left[A T^{\mu\nu}+BT g^{\mu\nu} \right] + C T^{\alpha\nu}\Gamma^{\prime\mu}_{\alpha\nu}= -\frac{g^{\mu\nu}}{4}\left(\frac{R}{\kappa} -  T \right)_{,\nu}.
\end{equation}
When $T^{\mu\nu}=0$ this gives $R=\text{constant}$ (with $\frac{\sigma_{,\nu}}{\sigma}$ undetermined).  Substituting Eq. (\ref{lambda}) into Eq. (\ref{sigmaequationb}) we find:
\begin{equation}
\label{newccconstant}
g^{\mu\nu}\frac{\partial \lambda}{\partial x^{\nu}}=2\frac{\sigma_{,\nu}}{\sigma}
\left[A\kappa T^{\mu\nu}+B\kappa T g^{\mu\nu} \right]+2C T^{\alpha\nu}\Gamma^{\prime\mu}_{\alpha\nu},
\end{equation}
which is equivalent to Eq. (\ref{constantcosmologicalconstant}) in the absence of matter.  We see that matter described by $T^{\mu\nu}$ can introduce a space-time dependence into the cosmological constant term without the need for dark energy or quintessence \cite{peeblesbharat}.  

In addition, Eq. (\ref{modconservation2b}) predicts mass creation and annihilation; this results from gradients in the measure and null-cone fields in regions where the energy-momentum tensor is non-zero.  If we consider a (cosmological) model where the only non-zero component of $T^{\mu\nu}$ is $T^{00}=c^2\rho$, where $\rho$ is the proper density of matter, then Eq. (\ref{modconservation2b}) yields (assuming a diagonal metric):
\begin{eqnarray}
\label{masscreation}
T^{0\nu}_{\;\; ;\nu}&=& \frac{\sigma_{,0}}{\sigma}\rho c^2\left[A-Bg^{00}  \right]+\frac{C\rho c^2}{2}\gamma^{00}\gamma_{00,0}\nonumber \\
T^{i\nu}_{\;\; ;\nu}&=& -B\rho c^2 g^{ii}\frac{\sigma_{,i}}{\sigma}-\frac{C\rho c^2}{2}\gamma^{ii}\gamma_{00,i}, 
\end{eqnarray}
where $i=1\ldots 3$ is not summed.  The coefficients $A$, $B$, and $C$ determine the net change in the energy-momentum density as the universe evolves.       

The equations of motion of a free dust particle follow from the matter conservation law (\ref{modconservation2b}).  Substituting the stress-energy tensor for dust $T^{\mu\nu}=\rho_0 u^{\mu}u^{\nu}$ into Eq. (\ref{modconservation2b}), where 
$u^{\mu}=\frac{dx^{\mu}}{d\tau}$ is the four-velocity (with $\tau$ proper time) that satisfies $u^{\mu}u_{\mu}=-c^2$ we obtain:
\begin{equation}
\label{modconservation3}
u^{\mu}u^{\nu}_{\; ;\nu}+u^{\nu}u^{\mu}_{\; ;\nu}=\frac{\sigma_{,\nu}}{\sigma}
\left[Au^{\mu}u^{\nu}-Bc^2 g^{\mu\nu} \right]+C u^{\alpha}u^{\nu}\Gamma^{\prime\mu}_{\alpha\nu}.
\end{equation}
Contracting with $u_{\mu}$ and noting $u_{\mu}u^{\mu}_{\; ;\nu}=0$ we find:
\begin{equation}
\label{modconservation4}
u^{\nu}_{\; ;\nu}=\left(A+B \right)\frac{\sigma_{,\nu}}{\sigma}u^{\nu}-\frac{C}{c^2} u_{\beta}u^{\alpha}u^{\nu}\Gamma^{\prime\beta}_{\alpha\nu}.
\end{equation}
We see that the momentum density $\rho_0u^{\mu}$ is conserved if $A=-B$ and $C=0$.  Substituting Eq. (\ref{modconservation4}) back into Eq. (\ref{modconservation3}) gives:
\begin{eqnarray}
\label{modconservation5}
u^{\nu}u^{\mu}_{\; ;\nu}&=&-B\frac{\sigma_{,\nu}}{\sigma}\left[u^{\mu}u^{\nu}+c^2g^{\mu\nu}\right]\nonumber\\
&+&Cu^{\alpha}u^{\nu}\left[\Gamma^{\prime\mu}_{\alpha\nu}+\frac{1}{c^2}\Gamma^{\prime\beta}_{\alpha\nu} u_{\beta}u^{\mu}   \right].
\end{eqnarray}
Consequently, the equations of motion for dust are:
\begin{eqnarray}
\label{geodesicequations}
\frac{d^2 x^\mu}{d\tau^2} &+& \Gamma^{\mu}_{\alpha\beta}\frac{dx^\alpha}{d\tau}\frac{dx^\beta}{d\tau}
=-B\frac{\sigma_{,\nu}}{\sigma}\left[\frac{dx^\mu}{d\tau}\frac{dx^\nu}{d\tau}
+c^2g^{\mu\nu}\right]\nonumber \\&+&C\frac{dx^\alpha}{d\tau}\frac{dx^\nu}{d\tau}\left[\Gamma^{\prime\mu}_{\alpha\nu}+\frac{1}{c^2}\Gamma^{\prime\beta}_{\alpha\nu} \frac{dx_\beta}{d\tau}\frac{dx^\mu}{d\tau}   \right].
\end{eqnarray}
We see that an additional force, proportional to the derivatives of the measure and null-cone fields, modifies the equations of motion of a free dust particle.  The equations of motion depend on the coefficients $B$ and $C$ explicitly and the coefficient $A$ implicitly through Eq. (\ref{sigmaequationb}).  If $B=C=0$ then they reduce to the usual geodesic equations of motion.  For the case $C=0$ the equations of motion follow from either the Lagrangian:
\begin{equation}
\label{lagrangian}
L=\frac{1}{2}\sigma^B g_{\mu\nu}\frac{dx^{\mu}}{d\tau}\frac{dx^{\nu}}{d\tau}-\frac{1}{2}\sigma^Bc^2,
\end{equation}
which is numerically equal to zero, or the Lagrangian:
\begin{equation}
\label{lagrangian2}
L=\sigma^B \sqrt{g_{\mu\nu}\frac{dx^{\mu}}{d\tau}\frac{dx^{\nu}}{d\tau}}.
\end{equation} 
  
\section{\label{sec:newtonianlimit}Newtonian Limit}
We investigate the Newtonian limit by considering the motion of a free dust particle in a circular orbit in the field of a spherically symmetric, static, localized mass of total gravitational mass $M$.  In addition, we assume that a spherically symmetric, static, matter density $\rho(r)$ is distributed in the space outside the localized source so that $T^{00}=c^2\rho(r)$ (with $T^{\mu\nu}=0$ for all other components).  We assume that the matter density outside the source $M$ does not contribute appreciably to the traditional gravitational force.  Thus, $M_{\rho}\ll M$, where $M_{\rho}$ is the total integrated mass contribution from $\rho(r)$ as observed from infinity.  This scenario will serve as a model for the motion of gas clouds in the disks of spiral galaxies at radii well beyond the visible disk's edge. 

In order to determine the radial force on a dust particle in a circular orbit we need to solve for $g_{\mu\nu}$ under the assumptions stated above.  Assuming $\sigma\simeq 1$, so that $\sigma = 1 + \epsilon(r)$, where $\epsilon(r) \ll 1$ we may write $g_{\mu\nu}\simeq g^{(M)}_{\mu\nu}$, where $g^{(M)}_{\mu\nu}$ is the solution of the field equations (\ref{traceless}) due to the localized mass $M$.  Since Eq. (\ref{traceless}) in the absence of matter is equivalent to the free-field Einstein field equations with an arbitrary cosmological constant, $g^{(M)}_{\mu\nu}$ is:
\begin{eqnarray}
\label{schwarzschild2}
g^{(M)}_{\mu\nu}=&-&\left(1-\frac{2m}{r}- \frac{\Lambda}{3}r^2\right)c^2dt^2+\frac{dr^2}{\left(1-\frac{2m}{r}- \frac{\Lambda}{3}r^2\right)} \nonumber \\&+&r^2\left(d\theta^2+\sin^2\theta d\phi^2 \right),
\end{eqnarray}
where $m=MG/c^2$ and $\Lambda=\frac{\lambda}{2}$ is the cosmological constant.      

Using Eqs.  (\ref{sigmaequationb}), (\ref{geodesicequations}), and (\ref{schwarzschild2}) we can determine the radial acceleration of gravity in the limit $\frac{dx^i}{d\tau}\ll c$:
\begin{eqnarray}
\label{gravityacceleration1}
g^{1}&=&\frac{d^2x^1}{d\tau^2}\simeq \frac{c^2}{2}g_{00,1}-Bc^2\frac{\sigma_{,1}}{\sigma}g^{11}+Cc^2\Gamma^{\prime 1}_{0 0}\nonumber \\
&=&-\frac{mc^2}{r^2}+\frac{\Lambda c^2}{3}r -\frac{c^2\rho_{,1}}{4\rho}- \frac{R_{,1}}{4\rho\kappa}.
\end{eqnarray}
We see that the gravitational acceleration is modified by two terms, one proportional to the gradient of the mass density, and another proportional to the gradient of the scalar curvature.  If we assume that the measure field is proportional to the matter density, then we may write (cf. Eq. (\ref{sigmaequationb})):
\begin{equation}
\label{scaleapprox}
R_{,1}=\alpha \kappa T_{,1}=-\alpha\kappa c^2\rho_{,1},
\end{equation}
where $\alpha$ is a constant.  In other words, matter determines scale only to the extent that the gradient in the curvature scalar is proportional to the gradient in mass density.  Consequently, Eq. (\ref{gravityacceleration1}) becomes:
\begin{equation}
\label{gravityacceleration2}
g^{1}=-\frac{mc^2}{r^2}+\frac{\Lambda c^2}{3}r + (\alpha-1)\frac{c^2}{4}\frac{\rho_{,1}}{\rho}.
\end{equation}
Note that $\rho_{,1}<0$ for any reasonable matter distribution, therefore, the sign of $(\alpha-1)$ determines whether or not the additional force is repulsive or attractive.  For the case $(\alpha-1)>0$ this additional acceleration may be responsible for the acceleration discrepancies observed on astrophysical scales, thus removing the need for dark matter.  Using Eq. (\ref{gravityacceleration2}), we can calculate the density distribution of visible matter that would produce a flat rotation curve beyond the visible edge of galaxies.  The velocity for a bound circular orbit is:
\begin{equation}
\label{flatrotcurve1}
v^2=\frac{mc^2}{r}-(\alpha-1)\frac{c^2}{4}\frac{\rho_{,1}}{\rho} r,
\end{equation}
where we have ignored the term due to the cosmological constant.  This will be constant as long as:
\begin{equation}
\label{flatrotcurve2}
(\alpha-1)\frac{c^2}{4}\frac{\rho_{,1}}{\rho} r=\frac{mc^2}{r}-v_0^2,
\end{equation}
where $v_0$ is the constant value of the rotation curve.
Solving the above differential equation yields:
\begin{equation}
\label{flatrotcurve3}
\rho(r)=\rho_0\exp\left[-\frac{4m}{(\alpha-1)r}\right]r^{-\beta},
\end{equation} 
where $\beta = \frac{4v_0^2}{(\alpha-1)c^2}$.  In the limit that the gravitational force $\frac{mc^2}{r^2}$ is negligible, the density decays as $r^{-\beta}$.  Assuming $\beta$ is on the order of unity and $v_0\sim 10^7 \;\text{cm}\cdot\text{s}^{-1}$, we estimate $(\alpha-1)\sim 10^{-7}$. 

In addition to the anomalous accelerations measured in the extragalactic regime, other evidence has been put forward in favor of dark matter.  For example, image distortion due to gravitational lensing for a large number of distant background galaxies suggests that $90\%$ of the matter of the foreground (lens) galaxies is invisible \cite{brainerdetal}.  For the case $C=0$ we may use either Lagrangian (\ref{lagrangian}) or (\ref{lagrangian2}) to obtain the equations of motion for massless particles:
\begin{equation}
\label{geodesicequations2}
\frac{d^2 x^\mu}{dp^2} + \Gamma^{\mu}_{\alpha\beta}\frac{dx^\alpha}{dp}\frac{dx^\beta}{dp}
=-B\frac{\sigma_{,\nu}}{\sigma}\left[\frac{dx^\mu}{dp}\frac{dx^\nu}{dp}\right],
\end{equation}
where $p$ is an affine parameter.  Hence, in order to compute the Einstein deflection angle one needs to know the mass density profile (cf. Eq. (\ref{sigmaequationb})) through which the photons propagate.  With the assumptions stated above, it is straightforward to show that the orbits of massless particles in a plane with polar coordinates $\{r=u^{-1},\phi\}$ around a point mass $m$ satisfy (cf. \cite{weinbergbook}):
\begin{equation}
\label{photonorbit}
\frac{d^2u}{d\phi^2}+u-3mu^2=\frac{B}{\sigma}\frac{d\sigma}{du}\left[u^2-\frac{1}{k^2}\right],
\end{equation}
where $k=\frac{\partial L}{\partial \left(\frac{d\phi}{dp}\right)}$.  This is equivalent to the standard case when $\frac{d\sigma}{du}=0$.  Therefore, mass determination via gravitational lensing will be inaccurate if the right hand side of Eq. (\ref{photonorbit}) is not included. 
  
As is well known, analyses of radio Doppler and ranging data from the Pioneer missions indicate that there is an apparent anomalous acceleration $a_0\sim 8\times 10^{-8} \text{cm}/\text{s}^2$ directed towards the sun \cite{andersonetal1988}.  According to the conclusions above, this acceleration may be attributed to the gradient in the mass density.  As before, we may calculate the density distribution that would produce the anomalous acceleration.  Thus, we find:
\begin{equation}
\label{Pioneer2}
\rho(r)=\rho_0\exp\left(-\frac{4a_0}{(\alpha-1)c^2}r\right),
\end{equation}
where $\rho_0$ is a constant.  If $(\alpha-1)\sim 10^{-7}$ as concluded above and $r\sim 10^{15} \;\text{cm}$, then $\rho$ would be roughly constant over this range of radii.  On the other hand, assuming the term in the exponential is on the order of unity, then $(\alpha-1)\sim 10^{-13}$.

%

\bibliography{urdm}

\end{document}